\documentclass{elsart}
\usepackage{graphicx}
\def\numt#1#2{#1 \times 10^{#2}}
\def\l{\left}
\def\r{\right}
\def\cU#1#2{U_{#1}^{#2}}

\def\Fgref#1{Fig.~\ref{fig:#1}}
\def\Fglab#1{\label{fig:#1}}
\def\gsim{~{\rlap{\lower 3.5pt\hbox{$\mathchar\sim$}}\raise 1pt\hbox{$>$}}\,}
\def\lsim{~{\rlap{\lower 3.5pt\hbox{$\mathchar\sim$}}\raise 1pt\hbox{$<$}}\,}
\def\PTP#1#2#3{Prog. Theor. Phys. {\bf #1}, #2 (#3)}
\def\PL#1#2#3{Phys. Lett. {\bf #1}, #2 (#3)}
\def\PRL#1#2#3{Phys. Pev. Lett. {\bf #1}, #2 (#3)}
\begin{document}
\begin{flushright}
VPI-IPPAP-02-04\\
hep-ph/0204118
\end{flushright}
\begin{frontmatter}
\title{
Possibility of the LBL experiment
with the high intensity proton
accelerator\thanksref{talk}
}
\thanks[talk]{
This talk is given at NuFACT'01 Workshop, Tsukuba, Japan, May 2001.
This paper is based on Ref.\cite{H2B}.
}
\author{N. Okamura}
\address{
Institute for particle Physics and Astrophysics,\\
Physics Departments, Virginia Tech,
Blacksburg, VA 24061
}
\ead{nokamura@vt.edu}
\begin{abstract}
We study physics possibility of Very Long
Base-Line (VLBL) Neutrino-Oscillation Experiments with the
High Intensity Proton Accelerator, which will be completed by the
year 2007 in Tokai-village, Japan.
As a target, a 100 kton-level water-$\check {\rm C}$erenkov detector
is considered at 2,100 km away.
Assuming the pulsed narrow-band $\nu_\mu$ beams, we study
sensitivity of such experiments to the neutrino mass hierarchy, 
the mass-squared differences,
one CP phase and three angles 
of the lepton-flavor-mixing matrix.
We find that experiments at a distance 2,100 km can determine 
the neutrino mass hierarchy if the mixing matrix element
$|U_{e3}|$ is not too small.
The CP phase and $|U_{e3}|$ can be constrained
if the large-mixing-angle solution of the solar-neutrino deficit
is realized.
\end{abstract}
\end{frontmatter}
In order to measure the neutrino oscillation parameters,
such as the neutrino mass-squared differences and 
the elements of the 3$\times$3 Maki-Nakagawa-Sakata (MNS) lepton
flavor-mixing matrix elements \cite{MNS}, 
various long base-line (LBL)
neutrino oscillation experiments are proposed.
In Japan, as a sequel to the K2K experiment, 
a new LBL neutrino oscillation experiment between the
High Intensity Proton Accelerator (HIPA) \cite{HIPA}
and the Super-Kamiokande (SK) 
with the base-line length of $L$=295 km 
has been proposed \cite{H2SK}.
In this talk, I discuss the
physics potential of Very Long Base-Line
(VLBL) neutrino oscillation experiments with HIPA and
a huge neutrino detector \cite{BAND}
in Beijing, at about $L$=2,100 km away.
As a beam option, 
the pulsed narrow-band $\nu^{}_{\mu}$ beam (NBB) is assumed.
For a target at $L$=2,100 km,
we consider a 100 kton water-$\check {\rm C}$erenkov detector
which is capable of 
measuring the both $\nu_\mu^{}$-like and $\nu_e^{}$-like events.
We study
the sensitivity of such experiments to the neutrino mass hierarchy,
the mass-squared differences, the three angles and one CP phase of the
MNS matrix \cite{H2B}.

In general, the MNS matrix has three mixing angles and three phases.
Two Majorana phases do not contribute to the neutrino oscillation.
Without losing generality, we can take 
$U_{e2}$ and $U_{\mu 3}$ to be real and non-negative.
By allowing $U_{e3}$ to have the complex phase 
$U_{e3}=|U_{e3}| e^{-i\delta_{_{\rm MNS}}} 
~(0 \leq \delta_{_{\rm MNS}} < 2\pi )\,,$
the four independent parameters are $U_{e2}, U_{\mu 3}, |U_{e3}|$
and $\delta_{_{\rm MNS}}$.
The constraints on the MNS matrix and the mass-squared differences
are given by the atmospheric-neutrino\cite{atm_SK},
solar-neutrino\cite{solar_SK}
and
the CHOOZ reactor experiments\cite{CHOOZ}. 
An analysis of the atmospheric-neutrino data
from the SK experiment \cite{atm_SK} finds
$\sin^2 2\theta_{_{\rm ATM}}\sim (0.88-1.0)$ and
$\delta m^2_{_{\rm ATM}} ({\rm eV}^2) 
\sim (1.6-4.0) \times 10^{-3}$.
From the observations of the solar-neutrino deficit by
the SK collaboration \cite{solar_SK}, 
the MSW large-mixing-angle solution (LMA) is
preferred to the others.
The CHOOZ experiment \cite{CHOOZ} 
gives the constraint
$\sin^2 2\theta_{_{\rm CHOOZ}}< 0.1$ for
$\delta m^2_{_{\rm CHOOZ}} > \numt{1.0}{-3}\,{\rm eV}^2$.
From these experiments,
the independent parameters in the MNS matrix
are obtained as
\begin{eqnarray}
\cU{\mu 3}{}&=&
\sqrt{1-\sqrt{1-\sin^2 2\theta_{_{\rm ATM}}}}\Big/\sqrt{2}\,,  \\
\cU{e2}{} &=&
\sqrt{1-|\cU{e3}{}|^2 -
\sqrt{\left(1-|\cU{e3}{}|^2\right)^2-\sin^22\theta_{_{\rm SOL}}}}
\Big/\sqrt{2}\,, \\
\l|\cU{e3}{}\r| &=& \sqrt{
1-\sqrt{1-\sin^2 2\theta_{_{\rm CHOOZ}}}
}\Big/\sqrt{2}\,.
\end{eqnarray}
All the other matrix elements
are then determined by the unitary conditions.
Hereafter,
we have made the identification
$
\delta m^2_{_{\rm SOL}} =
 \left|\delta m^2_{12}\right| 
\ll
 \left|\delta m^2_{13}\right| = \delta m^2_{_{\rm ATM}}\,,
$
with $\delta m^2_{ij}=m^2_j-m^2_i$. 

Since all the above constraints are obtained from
the survival probabilities which are even-functions
of $\delta m^2_{ij}$,
there are four neutrino-mass hierarchy cases corresponding to the
sign of the $\delta m^2_{ij}$ as shown in \Fgref{cases}.
If the MSW effect is relevant for
the solar-neutrino oscillation, then the hierarchy cases
II and IV are 
not favored.
The hierarchy I (III) is called `normal' (`inverted') hierarchy,
which corresponds to $\delta m^2_{12} > 0$ and 
$\delta m^2_{13} > 0$ ( $\delta m^2_{13} < 0$).
\begin{figure}[htb]
\begin{center}
\includegraphics[angle=-90,scale=0.5]{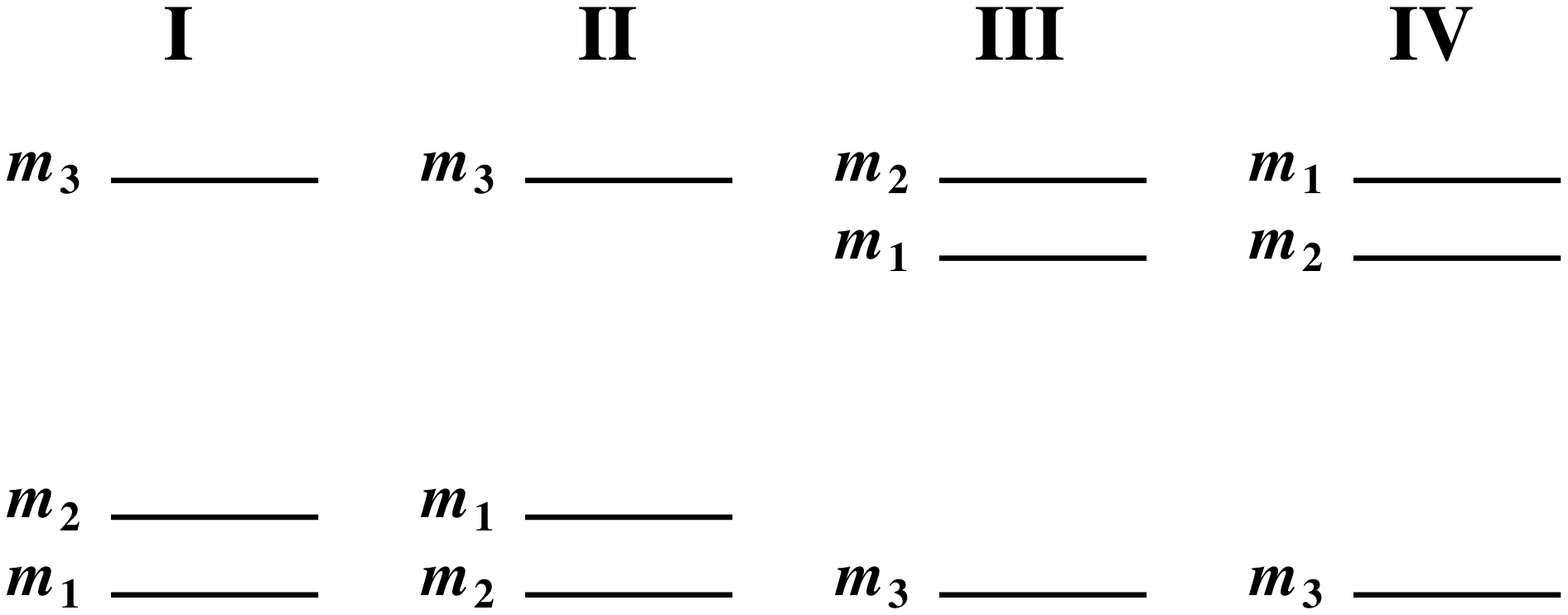}
\caption{Schematical view of the four cases of neutrino-mass hierarchy.}
\Fglab{cases}
\end{center}
\end{figure}

In order to examine the capability of the VLBL experiments
in determining the parameters,
we use the $\chi^2$ which is a function of the three angles, the two
mass-squared differences, the CP phase, the flux normalization factors
and the matter density.
The event number for the $\chi^2$
are derived from combining two experiments with different base-line length,
$L=2,100$ km (HIPA-to-Beijing) and $L=295$ km (HIPA-to-SK).
For a VLBL experiment at $L=2,100$ km,
we assume the statistical significance of
500 kton$\cdot$year each with two NBB whose peak energy is
4GeV and 6GeV respectively.
As for the LBL experiment at $L$= 295 km, we assume that 100
kton$\cdot$year for the low-energy NBB with $\langle p_\pi\rangle$= 2
GeV (NBB(2$\pi$)).
The data obtained from the LBL experiment is
what SK can gather in approximately 5 years with $10^{21}$ POT par year.

Thanks to the enhancement of matter effect, it is expected to 
distinguish the neutrino-mass hierarchy cases in such VLBL experiments.
We can determine the neutrino mass hierarchy at 3$\sigma$ level if 
$\sin^22\theta_{_{\rm CHOOZ}} >$0.04 by using this combination.
Also it is found that, if the LMA scenario is realized in Nature,
$\sin^2 2 \theta_{_{\rm CHOOZ}}$ and $\delta_{_{\rm MNS}}$ can be 
constrained at 1$\sigma$ level in favorable cases.
If the LMA scenario is realized in Nature and 
$\sin^22\theta_{_{\rm CHOOZ}}\gsim 0.06$, 
the SMA/LOW/VO scenarios 
can be rejected at 1$\sigma$ level 
when $\delta_{_{\rm MNS}}$ is around
$0^\circ$ or $180^\circ$ but not at all
when $\delta_{_{\rm MNS}}$ is around
$90^\circ$ or $270^\circ$.
For the atmospheric-neutrino oscillation parameters,
$\sin^22\theta_{_{\rm ATM}}$ is measured to 1$\%$ level and
$\delta m^2_{_{\rm ATM}}$ with the 3$\%$ accuracy
when $\sin^22\theta_{_{\rm ATM}}=1.0$,
$\delta m^2_{_{\rm ATM}}=3.5 \times 10^{-3}$ 
and the LMA scenario are realized in Nature.

\end{document}